\documentclass[12pt]{article}
\usepackage{bbm}
\newcommand{\be}{\begin{equation}}
\newcommand{\ee}{\end{equation}}
\newcommand{\ba}{\begin{eqnarray}}
\newcommand{\ea}{\end{eqnarray}}

\newcommand{\G}{\mathcal{G}}
\newcommand{\x}{\mbox{\boldmath$\times$}}
\begin{document}
\title{\normalsize \hfill UWThPh-2004-5
\\*[4mm] \LARGE
Symmetry realization of texture zeros}
\author{Walter Grimus\thanks{E-mail: walter.grimus@univie.ac.at} \\
\setcounter{footnote}{2}
\small Institut f\"ur Theoretische Physik, Universit\"at Wien \\
\small Boltzmanngasse 5, A--1090 Wien, Austria
\\*[3mm]
Anjan S.\ Joshipura\thanks{E-mail: anjan@prl.ernet.in} \\
\small Physical Research Laboratory, Ahmedabad 380009, India 
\\*[3mm]
Lu\'\i s Lavoura\thanks{E-mail: balio@cfif.ist.utl.pt} \\
\setcounter{footnote}{6}
\small Universidade T\'ecnica de Lisboa \\
\small Centro de F\'\i sica das Interac\c c\~oes Fundamentais \\
\small Instituto Superior T\'ecnico, P--1049-001 Lisboa, Portugal
\\*[3mm]
Morimitsu Tanimoto\thanks{E-mail: tanimoto@muse.sc.niigata-u.ac.jp} \\
\small Department of Physics, Niigata University \\
\small Ikarashi 2-8050, 950-2181 Niigata, Japan
\\*[4mm] 
}

\date{26 May 2004}

\maketitle

\begin{abstract}
We show that it is possible to enforce texture zeros
in arbitrary entries of the fermion mass matrices
by means of Abelian symmetries;
in this way,
many popular mass-matrix textures find a symmetry justification.
We propose two alternative methods which allow to place zeros
in any number of elements of the mass matrices that one wants.
They are applicable simultaneously in the quark and lepton sectors.
They are also applicable in Grand Unified Theories.
The number of scalar fields required by our methods may be large;
still,
in many interesting cases this number can be reduced considerably.
The larger the desired number of texture zeros is,
the simpler are the models which reproduce the texture.
\end{abstract}

\newpage

\section{Introduction}

The data from recent neutrino experiments \cite{SKam,SKamsolar,SNO,KamLAND}
confirm the neutrino oscillation hypothesis \cite{pontecorvo} and,
therefore,
confirm the hypothesis \cite{MNS} of neutrino masses
and lepton mixing---for recent reviews see \cite{reviews}.
While experiments have made great progress
in determining the lepton mixing matrix
and the two neutrino mass-squared differences,
the origin of neutrino masses and lepton mixing is still far from clear,
as witnessed by the many ideas put forward
in this context---see,
for instance,
\cite{models-reviews}.
Recently, Abelian symmetries have been studied systematically in
Ref.~\cite{low} with the aim of achieving extremal mixing angles.
A standard attempt at explaining the observed masses and mixing angles
is provided by `textures' for the lepton mass matrices,
with `texture zeros',
which have been widely discussed in the literature
\cite{FuNi,AK,Kang,Chen,desai}.
In particular,
it has been found that,
in the weak basis where the charged-lepton mass matrix is diagonal,
neutrino mass matrices with two texture zeros\footnote{Since
we are dealing with Majorana neutrinos,
the neutrino mass matrix $\mathcal{M}_\nu$ is symmetric
and $\left( \mathcal{M}_\nu \right)_{ab} =
\left( \mathcal{M}_\nu \right)_{ba} = 0$
is counted as only one texture zero when $a \neq b$.}
are consistent with all the experimental data \cite{frampton}.
The neutrino mass matrix does not seem to display
the hierarchical structure apparent in quark mass matrices
\cite{Xing1,Xing2,Barbieri,Ibara,HKT}.

The important theoretical problem with texture zeros is their origin.
In this paper,
we show that it is possible to enforce texture zeros
in arbitrary entries of the fermion mass matrices
by means of an Abelian symmetry.
We suggest two alternative general possibilities
for the Abelian symmetry group,\footnote{In the discovery
of the first alternative we have been inspired
by the `dimensional deconstruction'~\cite{hill}
model of~\cite{balaji}.}
which are simultaneously applicable in both the quark and lepton sectors,
and also in Grand Unified Theories.
They allow one to embed mass-matrix textures with zeros
into renormalizable field theories,
at the expense,
sometimes, of a proliferation of scalars and
flavour-changing neutral Yukawa interactions.

We explain the general methods in Section~\ref{method}
and illustrate them with four examples in Section~\ref{examples}.
Three examples are taken from the lepton sector
and use the seesaw mechanism \cite{seesaw}
to suppress the neutrino masses;
the last example is from the quark sector.
It is also shown in Section~\ref{examples} that,
in physically interesting cases,
the number of scalars needed to
implement a texture
can be reduced considerably relative to the general methods
of Section~\ref{method};
unfortunately,
it is not easy to provide a general way
for reducing the number of scalar multiplets,
or to find out the minimal number of scalars needed
in order to implement a given texture.
How to avoid potential Goldstone bosons
in our symmetry realization of texture zeros
is the subject theme of Section~\ref{goldstone}.
The conclusions are presented in Section~\ref{concl}.

\section{The methods}
\label{method}

We explain our methods by considering
the lepton sector of the Standard Model (SM),
which has three right-handed charged-lepton singlets $\ell_{Ra}$
and three left-handed lepton doublets $D_{La}$
($a = 1,2,3$);
to this we add three right-handed neutrino singlets $\nu_{Ra}$,
in order to enable the seesaw mechanism
for suppressing the neutrino masses.

\subsection{First method}

We assume that each of the nine fermion multiplets $f$
transforms under a \emph{separate}
Abelian group $\G \left( f \right)$,
so that the full horizontal symmetry group $\G$
is the direct product of nine groups:
\be
\G = \times_f\, \G \left( f \right),
\quad \mbox{with} \ f = \ell_{Ra}, D_{La}, \nu_{Ra} \ (a = 1,2,3).
\ee
For mathematical transparency
we identify the groups $\G \left( f \right)$
with their unitary representations;
thus,
the $\G \left( f \right)$ are groups of complex numbers on the unit circle,
either the discrete groups $\mathbbm{Z}_n$ or the continuous group $U(1)$.
We introduce Higgs doublets $\phi_{ab}$ and $\tilde\phi_{ab}$
(note that in general the $\phi_{ab}$ and the $\tilde\phi_{ab}$
are \emph{independent} degrees of freedom),
with weak hypercharges $+1/2$ and $-1/2$,
respectively,
together with scalar singlets $\chi_{ab}$ with weak hypercharge $0$.
They transform in the following way:
\be
\begin{array}{rl}
\phi_{ab}:  & \G^* \left( \ell_{Ra} \right)
\otimes \G \left( D_{Lb} \right), \\
\tilde\phi_{ab}:  & \G^* \left( \nu_{Ra} \right)
\otimes \G \left( D_{Lb} \right), \\
\chi_{ab}: & \G \left( \nu_{Ra} \right)
\otimes \G \left( \nu_{Rb} \right).
\end{array}
\label{scalarreps}
\ee
We make the identification $\chi_{ab} \equiv \chi_{ba}$,
$\forall\, a, b \in \left\{ 1, 2, 3 \right\}$,
so that there are only six
(in general complex)
scalar singlets. 
The $\G$-invariant Yukawa Lagrangian of the leptons is
\be
\mathcal{L}_{\mathrm{Y} \ell} = 
- \sum_{a,b} \left( 
\Gamma_{ab}\, \bar \ell_{Ra} \phi_{ab}^\dagger D_{Lb}
+ \Delta_{ab}\, \bar \nu_{Ra} {\tilde\phi}_{ab}^\dagger D_{Lb}
+ \frac{1}{2}\, Y_{ab}\, \chi_{ab} \bar \nu_{Ra} C \bar \nu_{Rb}^T
\right) + \mbox{H.c.}
\label{Lyukawa}
\ee
There is one coupling constant $\Gamma_{ab}$
for each Higgs doublet $\phi_{ab}$,
one coupling constant $\Delta_{ab}$
for each Higgs doublet $\tilde\phi_{ab}$,
and one coupling constant $Y_{ab}$
for each scalar singlet $\chi_{ab}$.
With the vacuum expectation values (VEVs) 
\be
\langle 0 | \phi_{ab} | 0 \rangle
= \left( \begin{array}{c} 0 \\ v_{ab} \end{array} \right), 
\quad
\langle 0 | \tilde\phi_{ab} | 0 \rangle
= \left( \begin{array}{c} w^*_{ab} \\ 0 \end{array} \right),
\quad
\left\langle 0 \left| \chi_{ab} \right| 0 \right\rangle = X_{ab},
\ee
one obtains the charged-lepton mass matrix $M_\ell$,
the neutrino Dirac mass matrix $M_D$,
and the right-handed-neutrino Majorana mass matrix $M_R$ through
\be
\left( M_\ell \right)_{ab} = v_{ab}^\ast \Gamma_{ab},
\quad
\left( M_D \right)_{ab}= w_{ab} \Delta_{ab},
\quad \mbox{and} \quad
\left( M_R \right)_{ab} = X_{ab} Y_{ab},
\label{MlD}
\ee
respectively.
It is natural to assume that the VEVs $X_{ab}$ of the scalar singlets
are of a very large (seesaw) scale.
One then obtains a suppressed Majorana mass matrix \cite{seesaw}
\be
\mathcal{M}_\nu = - M_D^T M_R^{-1} M_D
\label{seesaw}
\ee
for the light neutrinos.

In this scheme,
according to Eqs.~(\ref{scalarreps})--(\ref{MlD}),
for each non-zero entry in $M_\ell$ or $M_D$ we need one Higgs doublet,
with appropriate transformation properties under $\G$,
connecting the two fermion multiplets corresponding to that entry.
Similarly,
for each non-zero matrix element of $M_R$ we need a singlet $\chi_{ab}$
with appropriate transformation properties under $\G$.
Under these conditions,
it is easy to place a texture zero
in any entry of the mass matrices $M_\ell$,
$M_D$,
or $M_R$,
simply by not introducing into the theory
the corresponding scalar multiplet.
Thus,
if we want to have $\left( M_\ell \right)_{ab} \neq 0$,
then we allow the presence of $\phi_{ab}$ in the theory;
if,
on the contrary,
we want $\left( M_\ell \right)_{ab}$ to vanish,
then we do not admit the Higgs doublet $\phi_{ab}$
into the set of scalar fields.

For each of the matrices $M_\ell$ and $M_D$
we need at most nine Higgs doublets;
in the case of nine Higgs doublets we obtain
the most general mass matrix.
For a fully general $M_R$,
which is symmetric,
we need six scalar singlets.
In practice,
however,
in order to obtain predictive mass matrices
there will be many zero entries in those matrices,
and the actual number of scalar multiplets
will be much less than $9+9+6 = 24$.

If the horizontal-symmetry group corresponding to $\nu_{Ra}$
is $\mathbbm{Z}_2 \left( \nu_{Ra} \right)$,
then we obtain a non-zero
diagonal entry $\left (M_R \right)_{aa}$ in $M_R$
from the automatic presence in the Lagrangian of the mass term
$- m_a \bar \nu_{Ra} C \bar \nu_{Ra}^T + \mathrm{H.c.}$,
without the need for a singlet $\chi_{aa}$.
Thus,
if one wants to enforce $\left( M_R \right)_{aa} = 0$,
then $\nu_{Ra}$ must not transform as 
$\mathbbm{Z}_2 \left( \nu_{Ra} \right)$,
rather $\G \left( \nu_{Ra} \right)$ has to be genuinely complex.

In order to obtain a non-singular $M_\ell$ we need
at least three non-zero elements in that matrix and,
by adequately interchanging the $\ell_{Ra}$,
those three non-zero matrix elements
may be placed along the diagonal of $M_\ell$.
The number of Higgs doublets $\phi_{ab}$ needed
may then be reduced by assuming $\G  \left( \ell_{Ra} \right)
\equiv \G \left( D_{La} \right)$
and by using only one $\phi$ doublet,
invariant under the horizontal-symmetry group,
which simultaneously generates all three diagonal matrix elements
in $M_\ell$.
A similar trick may be used in $M_D$,
and also in $M_R$ if one wants to reduce the number of scalar singlets
employed.

If the zeros in $M_\ell$ and $M_D$ are at the same places,
and if the Higgs doublets responsible for $M_\ell$
transform under the same \emph{real} representation $\G$
as those responsible for $M_D$,
then we may identify $\tilde\phi_{ab}$
with $i \sigma_2 \phi_{ab}^*$;
in that case we have $\G \left( \ell_{Ra} \right)
\equiv \G \left( \nu_{Ra} \right)$,
$\G$ is the direct product of only six groups $\G \left( f \right)$,
and the number of Higgs doublets needed may be reduced considerably.

All these points will become clearer
after the examples in Section~\ref{examples}.

\subsection{Second method}

An alternative and much simpler possibility
for the Abelian symmetry group is the general choice
\be
\G  = \mathbbm{Z}_{12} \times \mathbbm{Z}_2.
\label{Z12}
\ee
Let $\omega = \exp \left( i \pi / 6 \right)$.
Under the $\mathbbm{Z}_{12}$ of Eq.~(\ref{Z12}),
\be
\begin{array}{rclcrclcrcl}
\bar \ell_{R1} &\to& \omega\, \bar \ell_{R1}, & &
\bar \nu_{R1} &\to& \omega\, \bar \nu_{R1}, & &
D_{L1} &\to& \omega\, D_{L1}, \\
\bar \ell_{R2} &\to& \omega^2 \bar \ell_{R2}, & &
\bar \nu_{R2} &\to& \omega^2 \bar \nu_{R2}, & &
D_{L2} &\to& \omega^3 D_{L2}, \\
\bar \ell_{R3} &\to& \omega^5 \bar \ell_{R3},  & &
\bar \nu_{R3} &\to& \omega^5 \bar \nu_{R3}, & &
D_{L3} &\to& \omega^8 D_{L3}.
\end{array}
\label{z12}
\ee
Then,
the bilinears $\bar \ell_{Ra} D_{Lb}$ and $\bar \nu_{Ra} D_{Lb}$,
relevant for $\left( M_\ell \right)_{ab}$ and $\left( M_D \right)_{ab}$,
respectively,
transform according to the matrix
\be
\left( \begin{array}{ccc}
\omega^2 & \omega^4 & \omega^9 \\
\omega^3 & \omega^5 & \omega^{10} \\
\omega^6 & \omega^8 & \omega
\end{array} \right),
\label{first}
\ee
while the bilinears $\bar \nu_{Ra} C \bar \nu_{Rb}^T$,
relevant for $\left( M_R \right)_{ab}$,
transform according to
\be
\left( \begin{array}{ccc}
\omega^2 & \omega^3 & \omega^6 \\
\omega^3 & \omega^4 & \omega^7 \\
\omega^6 & \omega^7 & \omega^{10}
\end{array} \right).
\label{second}
\ee
Since all the powers of $\omega$
in the matrix of~(\ref{second}) are different,
we may introduce into the theory singlets $\chi_{ab}$
with the appropriate transformation properties under $\mathbbm{Z}_{12}$
in order to render non-zero only those matrix elements of $M_R$
that one wants;
notice that no bilinear $\bar \nu_{Ra} C \bar \nu_{Rb}^T$
is $\G$-invariant,
and therefore $\chi_{ab}$ is always needed
in order to obtain $\left( M_R \right)_{ab} \neq 0$.
Similarly,
all powers of $\omega$
in the matrix of~(\ref{first}) are different,
and therefore one needs a separate $\phi_{ab}$ or $\tilde \phi_{ab}$
in order to make each matrix element $\left( M_\ell \right)_{ab}$
or $\left( M_D \right)_{ab}$,
respectively,
non-zero.

The factor group $\mathbbm{Z}_2$ in Eq.~(\ref{Z12})
is needed in order to distinguish the $\phi_{ab}$
from the $\tilde \phi_{ab}$,
so that some Higgs doublets do not simultaneously generate
non-zero matrix elements in $M_\ell$ and in $M_D$.
Under that $\mathbbm{Z}_2$ the $\tilde \phi_{ab}$
and the neutrino singlets $\nu_{Ra}$ change sign,
while all other multiplets remain invariant.

Variations on this second method are,
of course,
possible.
One may modify the transformation properties of the $D_{La}$
by a fixed power of $\omega$,
for instance.
Or one may substitute $\mathbbm{Z}_{12}$ by a $U(1)$ group,
by trading the $\omega^q$ in Eq.~(\ref{z12}) by $e^{i q \alpha}$,
with $\alpha$ a continuous real parameter.
Most important,
in the realization of many specific textures
one may trade the symmetry $\mathbbm{Z}_{12} \times \mathbbm{Z}_2$
by a smaller symmetry;
we will encounter examples with $\mathbbm{Z}_8$ in the following section.

\section{Examples}
\label{examples}

We have borrowed the first two examples from~\cite{honda},
where they are called cases $b_2$ and $a_1$,
respectively;
they reproduce the texture $A_2$ of~\cite{frampton}
for $\mathcal{M}_\nu$:
\be
\mathcal{M}_\nu \sim 
\left( \begin{array}{ccc}
0 & \x & 0 \\ \x & \x & \x \\ 0 & \x & \x 
\end{array} \right),
\ee
where the crosses represent non-zero matrix elements.
In both examples $M_\ell$ is diagonal,
hence we may identify the indices $1,2,3$ with $e,\mu,\tau$,
respectively.
\paragraph{Example A:}
\be
M_\ell\ \mbox{and}\ M_D\ \mbox{diagonal}, \quad
M_R \sim \left( \begin{array}{ccc}
0 & \x & \x \\ \x & 0 & 0 \\ \x & 0 & \x \end{array} \right).
\ee
This is a particularly simple case
since $M_\ell$ and $M_D$ are simultaneously diagonal.
Using the first method of the previous section,
this allows the identification
$\G \left( \ell_{Ra} \right) \equiv \G \left( \nu_{Ra} \right)
\equiv \G \left( D_{La} \right)$,
so that $\G$ is the direct product of only three groups.
One Higgs doublet $\phi$ transforming trivially under $\G$,
together with $\tilde\phi = i\sigma_2 \phi^*$,
are sufficient for generating both $M_\ell$ and $M_D$.
Since $\left( M_R \right)_{11} = \left( M_R \right)_{22} = 0$,
we take $\G \left( \nu_{R1} \right)
\equiv \mathbbm{Z}_4 \left( \nu_{R1} \right)$
and $\G \left( \nu_{R2} \right)
\equiv \mathbbm{Z}_4 \left( \nu_{R2} \right)$;
for $\G \left( \nu_{R3} \right)$
we use $\G \left( \nu_{R3} \right)
\equiv \mathbbm{Z}_2 \left( \nu_{R3} \right)$
so that $\left( M_R \right)_{33}$ is non-zero
even in the absence of any scalar singlets.
We need two scalar singlets,
transforming as
\be
\begin{array}{rl}
\chi_{12}: & \mathbbm{Z}_4 \left( \nu_{R1} \right)
\otimes \mathbbm{Z}_4 \left( \nu_{R2} \right), \\
\chi_{13}: & \mathbbm{Z}_4 \left( \nu_{R1} \right)
\otimes \mathbbm{Z}_2 \left( \nu_{R3} \right).
\end{array}
\label{Atransf}
\ee
Thus,
in this example there is only the SM Higgs doublet,
together with two complex scalar singlets
which at low energies are invisible. 
\paragraph{Example B:} 
\be
M_\ell\ \mbox{diagonal}, \quad
M_D \sim \left( \begin{array}{ccc}
\x & 0 & \x \\ 0 & 0 & \x \\ 0 & \x & 0 \end{array} \right), \quad
M_R \sim \left( \begin{array}{ccc}
\x & 0 & \x \\ 0 & \x & 0 \\ \x & 0 & 0 \end{array} \right).
\label{B}
\ee
Firstly,
we discuss a straightforward realization of this texture,
using the first method of Section~\ref{method}.
Since $M_\ell$ is diagonal,
we choose $\G \left( \ell_{Ra} \right)
\equiv \G \left( D_{La} \right)$,
and $\G$ is the direct product of only six groups.
In order to obtain the diagonal $M_\ell$,
it is then sufficient to
introduce one Higgs doublet $\phi$, 
transforming trivially under $\G$.
We further choose $\G \left( D_{La} \right)
= \mathbbm{Z}_2 \left( D_{La} \right)$,
$\forall \, a=1,2,3$.
Since $M_R$ has one zero in the diagonal,
we take a genuinely complex group for all the $\nu_{Ra}$,
viz.\ $\G \left( \nu_{Ra} \right)
= \mathbbm{Z}_4 \left( \nu_{Ra} \right)$. 
There are four non-zero entries in $M_D$,
therefore we need (besides $\phi$) four more Higgs doublets, 
with the following transformation properties:
\be
\begin{array}{rl}
\tilde\phi_{11}: & \mathbbm{Z}^*_4 \left( \nu_{R1} \right)
\otimes \mathbbm{Z}_2 \left( D_{L1} \right), \\
\tilde\phi_{13}: & \mathbbm{Z}^*_4 \left( \nu_{R1} \right)
\otimes \mathbbm{Z}_2 \left( D_{L3} \right), \\
\tilde\phi_{23}: & \mathbbm{Z}^*_4 \left( \nu_{R2} \right)
\otimes \mathbbm{Z}_2 \left( D_{L3} \right), \\
\tilde\phi_{32}: & \mathbbm{Z}^*_4 \left( \nu_{R3} \right)
\otimes \mathbbm{Z}_2 \left( D_{L2} \right).
\end{array}
\ee
It remains to discuss the scalar singlets:
\be
\begin{array}{rl}
\chi_{11}: & \mathbbm{Z}_4 \left( \nu_{R1} \right)
\otimes \mathbbm{Z}_4 \left( \nu_{R1} \right), \\
\chi_{22}: & \mathbbm{Z}_4 \left( \nu_{R2} \right)
\otimes \mathbbm{Z}_4 \left( \nu_{R2} \right), \\
\chi_{13}: & \mathbbm{Z}_4 \left( \nu_{R1} \right)
\otimes \mathbbm{Z}_4 \left( \nu_{R3} \right).
\end{array}
\ee
There are five Higgs doublets and three scalar singlets
in this straightforward realization of the texture.

Secondly,
one may try,
still using the first method of Section~\ref{method},
to reduce the number of scalars.
We stick to $\G \left( \ell_{Ra} \right)
\equiv \G \left( D_{La} \right)$
and use for $\G$ the direct product of five groups instead of six: 
\be
\begin{array}{l}
\mathbbm{Z}_2 \left( D_{L1} \right)
\equiv \mathbbm{Z}_2 \left( \ell_{R1} \right), \\
\mathbbm{Z}_4 \left( D_{L2} \right)
\equiv \mathbbm{Z}_4 \left( \ell_{R2} \right)
\equiv \mathbbm{Z}_4 \left( \nu_{R3} \right), \\
\mathbbm{Z}_2 \left( D_{L3} \right)
\equiv \mathbbm{Z}_2 \left( \ell_{R3} \right), \\
\mathbbm{Z}_2 \left( \nu_{R1} \right), \\
\mathbbm{Z}_2 \left( \nu_{R2} \right).
\end{array}
\ee
As before,
we introduce a Higgs doublet $\phi$ transforming trivially under $\G$.
Then the coupling of $\tilde \phi = i \sigma_2 \phi^\ast$
is responsible for $\left( M_D \right)_{32} \neq 0$.
For the three remaining non-zero elements of $M_D$
we need the following Higgs doublets:
\be
\begin{array}{rl}
\tilde \phi_{11}: & \mathbbm{Z}_2 \left( \nu_{R1} \right)
\otimes \mathbbm{Z}_2 \left( D_{L1} \right), \\
\tilde \phi_{13}: & \mathbbm{Z}_2 \left( \nu_{R1} \right)
\otimes \mathbbm{Z}_2 \left( D_{L3} \right), \\
\tilde \phi_{23}: & \mathbbm{Z}_2 \left( \nu_{R2} \right)
\otimes \mathbbm{Z}_2 \left( D_{L3} \right).
\end{array}
\ee
Now one scalar singlet $\chi_{13}$
transforming as $\mathbbm{Z}_2 \left( \nu_{R1} \right)
\otimes \mathbbm{Z}_4 \left( \nu_{R3} \right)$ is sufficient.
Thus,
in this realization of the texture of Eq.~(\ref{B})
we manage to have only four Higgs doublets and one scalar singlet,
while $\G$ is the direct product of four $\mathbbm{Z}_2$
and one $\mathbbm{Z}_4$ groups.

Thirdly,
we note that there are even more economic realizations
of the present texture,
when one uses the second method of the previous section,
or a simplified version thereof.
Consider for instance the following $\mathbbm{Z}_8$ symmetry:
\be
\begin{array}{rclcrclcrcl}
\bar \ell_{R2} &\to& \zeta \bar \ell_{R2}, & &
\bar \nu_{R2} &\to& \zeta^6 \bar \nu_{R2}, & &
D_{L2} &\to& \zeta^7 D_{L2},  \\
\bar \ell_{R3} &\to& \zeta^4 \bar \ell_{R3}, & &
\bar \nu_{R3} &\to& \zeta^3 \bar \nu_{R3}, & &
D_{L3} &\to& \zeta^2 D_{L3}, 
\end{array}
\ee
where $\zeta = \exp \left( i \pi / 4 \right)$.
We then need only two Higgs doublets $\phi$ and $\phi^\prime$,
one complex scalar singlet $\chi$,
and one real scalar singlet $\chi^\prime$,
transforming as
\be
\begin{array}{rcl}
\phi &\to& \phi, \\
\phi^\prime &\to& \zeta^6 \phi^\prime, \\
\chi &\to& \zeta^5 \chi, \\
\chi^\prime &\to& \zeta^4 \chi^\prime.
\end{array}
\ee
The Yukawa couplings matrices of $\phi$,
$\phi'$,
$\tilde \phi = i \sigma_2 \phi^\ast$,
and $\tilde \phi^\prime = i \sigma_2 {\phi^\prime}^\ast$ are given by 
\be
\Gamma \sim \left( \begin{array}{ccc} 
\x & 0 & 0 \\ 0 & \x & 0 \\ 0 & 0 & 0 \end{array} \right),
\ \
\Gamma' \sim \left( \begin{array}{ccc} 
0 & 0 & 0 \\ 0 & 0 & 0 \\ 0 & 0 & \x \end{array} \right),
\ \
\Delta \sim \left( \begin{array}{ccc} 
\x & 0 & 0 \\ 0 & 0 & \x \\ 0 & 0 & 0 \end{array} \right),
\ \
\Delta' \sim \left( \begin{array}{ccc} 
0 & 0 & \x \\ 0 & 0 & 0 \\ 0 & \x & 0 \end{array} \right),
\ee
respectively.
While the non-zero entry $\left( M_R \right)_{11}$ in the mass matrix
of the right-handed neutrino singlets does not need a scalar singlet,
the rest of the entries in $M_R$ of Eq.~(\ref{B}) is supplied by the
Yukawa coupling matrices 
\be
Y \sim \left( \begin{array}{ccc} 
0 & 0 & \x \\ 0 & 0 & 0 \\ \x & 0 & 0 \end{array} \right)
\quad \mbox{and} \quad 
Y' \sim \left( \begin{array}{ccc} 
0 & 0 & 0 \\ 0 & \x & 0 \\ 0 & 0 & 0 \end{array} \right)
\ee
of $\chi$ and $\chi'$, respectively.

Our second and third realizations of the texture in Eq.~(\ref{B})
illustrate the fact that the methods
presented in Section~\ref{method}
do not necessarily lead to the simplest Abelian symmetry
which justifies each texture,
nor to the minimal number of scalar multiplets or fields.
For some textures it may be possible to find simpler symmetries
than the ones presented in Section~\ref{method},
and realizations of the texture which require fewer scalar multiplets.

\paragraph{Example C:}
\be
M_\ell \sim \left( \begin{array}{ccc}
0 & \x & 0 \\ \x & 0 & \x \\ 0 & \x & \x \end{array} \right),
\quad
M_D \sim \left( \begin{array}{ccc}
0 & \x & 0 \\ \x & 0 & \x \\ 0 & \x & \x \end{array} \right),
\quad
M_R\ \mbox{diagonal}.
\label{C}
\ee
We have borrowed this example
from~\cite{fukugita} (see also~\cite{xing}),
where it is moreover assumed
that the diagonal elements of $M_R$ are all equal;
unfortunately,
the latter feature cannot be implemented
with the methods proposed in the present paper.
In $M_\ell$ and $M_D$ the position of the zeros is the same;
using the first method in Section~\ref{method},
this makes it convenient to choose $\G \left( \ell_{Ra} \right)
\equiv \G \left( \nu_{Ra} \right)$.
We set $\G \left( \nu_{Ra} \right)
= \mathbbm{Z}_2 \left( \nu_{Ra} \right)$,
an option which automatically yields a mass matrix $M_R$
with non-zero diagonal elements,  
without the need for any scalar singlets.
We choose $\G$ as being the direct product
of four $\mathbbm{Z}_2$ groups:
\ba
& & \mathbbm{Z}_2 \left( D_{L1} \right) \equiv
\mathbbm{Z}_2 \left( \ell_{R2} \right) \equiv
\mathbbm{Z}_2 \left( \nu_{R2} \right),
\label{Z21} \\
& & \mathbbm{Z}_2 \left( D_{L2} \right) \equiv
\mathbbm{Z}_2 \left( \ell_{R1} \right) \equiv
\mathbbm{Z}_2 \left( \nu_{R1} \right),
\label{Z22} \\
& &
\mathbbm{Z}_2 \left( \ell_{R3} \right) \equiv
\mathbbm{Z}_2 \left( \nu_{R3} \right),
\label{Z23} \\
& & \mathbbm{Z}_2 \left( D_{L3} \right).
\label{Z24}
\ea
Then we need four Higgs doublets:
$\phi_{12} \equiv \phi_{21}$,
which is invariant under $\G$;
$\phi_{33}$,
which changes sign under the $\mathbbm{Z}_2$ groups
in Eqs.~(\ref{Z23}) and~(\ref{Z24});
$\phi_{23}$,
which changes sign under the $\mathbbm{Z}_2$ groups
in Eqs.~(\ref{Z21}) and~(\ref{Z24});
and $\phi_{32}$,
which changes sign under the $\mathbbm{Z}_2$ groups
in Eqs.~(\ref{Z22}) and~(\ref{Z23}).
For all these Higgs doublets 
we need to consider the Yukawa couplings
of both the $\phi_{ab}$
and the $\tilde \phi_{ab} = i \sigma_2 \phi_{ab}^\ast$.
Thus,
we  reproduce the texture in Eq.~(\ref{C})
by using a horizontal symmetry group $\G$
which is the direct product of four $\mathbbm{Z}_2$ groups,
with a scalar sector consisting of four Higgs doublets.

Lastly,
we consider an example from the quark sector.
In that sector,
the quark doublets $q_{La}$,
singlets $u_{Ra}$ with charge $2/3$,
and singlets $d_{Ra}$ with charge $-1/3$
correspond to the $D_{La}$,
$\nu_{Ra}$,
and $\ell_{Ra}$,
respectively,
in the lepton sector.
Moreover,
the third term is missing
from the Yukawa Lagrangian in Eq.~(\ref{Lyukawa}).
\paragraph{Example D:}
\be
M_d \sim \left( \begin{array}{ccc}
0 & \x & 0 \\ \x & \x & \x \\ 0 & \x & \x \end{array} \right),
\quad
M_u \sim \left( \begin{array}{ccc}
0 & \x & 0 \\ \x & \x & \x \\ 0 & \x & \x \end{array} \right).
\label{D}
\ee
We have taken this texture from~\cite{roberts,fritzsch}
(see also the references therein).
Non-symmetric mass matrices are considered
in one of the instances in~\cite{roberts};
using the methods in the present paper,
it is possible
neither to achieve symmetric mass matrices $M_u$ and $M_d$,
nor Hermitian ones like those in~\cite{fritzsch}.
Using the first method in Section~\ref{method},
since we want zeros in the same positions of $M_u$ and $M_d$,
we take $\G \left( u_{Ra} \right) \equiv
\G \left( d_{Ra} \right)$.
The symmetry group $\G$ may be chosen
to be the direct product of
\be
\begin{array}{l}
\mathbbm{Z}_2 \left( q_{L1} \right), \\
\mathbbm{Z}_2 \left( d_{R1} \right) \equiv
\mathbbm{Z}_2 \left( u_{R1} \right), \\
\mathbbm{Z}_2 \left( q_{L2} \right)
\equiv \mathbbm{Z}_2 \left( d_{R2} \right)
\equiv \mathbbm{Z}_2 \left( u_{R2} \right), \\
\mathbbm{Z}_2 \left( q_{L3} \right)
\equiv \mathbbm{Z}_2 \left( d_{R3} \right)
\equiv \mathbbm{Z}_2 \left( u_{R3} \right).
\end{array}
\ee
Then,
with four Higgs doublets transforming as
\be
\begin{array}{rl}
\phi_{12}: & \mathbbm{Z}_2 \left( d_{R1} \right)
\otimes \mathbbm{Z}_2 \left( q_{L2} \right), \\
\phi_{21}: & \mathbbm{Z}_2 \left( d_{R2} \right)
\otimes \mathbbm{Z}_2 \left( q_{L1} \right), \\
\phi_{23} = \phi_{32}: & \mathbbm{Z}_2 \left( d_{R2} \right)
\otimes \mathbbm{Z}_2 \left( q_{L3} \right), \\
\phi_{22} = \phi_{33}: & \{ 1\},
\end{array}
\label{Dphi}
\ee
where $\{ 1\}$ denotes invariance under $\G$,
one reproduces the texture in Eq.~(\ref{D}).

If one adopts a variation of the second method in Section~\ref{method},
one can find a realization of the present texture
with only three Higgs doublets.
Using once again a $\mathbbm{Z}_8$ symmetry,
take
\be
\begin{array}{rclcrclcrcl}
\bar d_{R1} &\to& \bar d_{R1}, & &
\bar u_{R1} &\to& \bar u_{R1}, & &
q_{L1} &\to& \zeta^3 q_{L1}, \\
\bar d_{R2} &\to& \zeta^6 \bar d_{R2}, & &
\bar u_{R2} &\to& \zeta^6 \bar u_{R2}, & &
q_{L2} &\to& \zeta\, q_{L2}, \\ 
\bar d_{R3} &\to& \zeta^3 \bar d_{R3}, & &
\bar u_{R3} &\to& \zeta^3 \bar u_{R3}, & &
q_{L3} &\to& \zeta^6 q_{L3}.
\end{array}
\ee
One then needs only three Higgs doublets $\phi_1$,
$\phi_2$,
and $\phi_3$ transforming with $\zeta$,
$\zeta^7$,
and $\zeta^4 = -1$,
respectively,
in order to reproduce
the texture in Eq.~(\ref{D}).
The Yukawa coupling matrices are
\be
\Gamma_1 \sim \Delta_2 \sim 
\left( \begin{array}{ccc} 
0 & \x & 0 \\ \x & 0 & 0 \\ 0 & 0 & \x \end{array} \right),
\quad
\Gamma_2 \sim \Delta_1 \sim 
\left( \begin{array}{ccc} 
0 & 0 & 0 \\ 0 & \x & 0 \\ 0 & 0 & 0 \end{array} \right),
\quad 
\Gamma_3 \sim \Delta_3 \sim 
\left( \begin{array}{ccc} 
0 & 0 & 0 \\ 0 & 0 & \x \\ 0 & \x & 0 \end{array} \right).
\ee

\section{Goldstone bosons}
\label{goldstone}

The Abelian symmetry group $\G$ is often so restrictive
that it leads to accidental $U(1)$ symmetries in the scalar potential.
Those $U(1)$ symmetries
may either be shared by the rest of the Lagrangian,
or not.
Thus,
there is the danger of Goldstone or pseudo-Goldstone bosons. 
If a Goldstone boson is a superposition of scalar singlets,
or components thereof,
then it is harmless,
since it couples only to the right-handed neutrino singlets \cite{peccei}. 

In some cases there are no Goldstone bosons.
For instance,
if all scalars transform with $\mathbbm{Z}_2$ groups,
like for instance
in example C of Section~\ref{examples},
then the terms $\left( \phi_i^\dagger \phi_j \right)^2$ with $i \neq j$
are $\G$-invariant and eliminate all the Goldstone bosons.

It is always possible to avoid the Goldstone bosons
by breaking $\G$ softly in the scalar potential
through terms of dimension two.
We simply add to the potential all terms
of the form $\phi_i^\dagger \phi_j$ with $i \neq j$.
The only surviving $U(1)$ is then associated with hypercharge,
and no Goldstone boson coupling to $\bar \ell_{Ra} D_{Lb}$ occurs.
(If desirable,
one may also eliminate the Goldstone bosons
in the scalar-singlet sector in an analogous way.)
For instance,
our $\mathbbm{Z}_{12} \times \mathbbm{Z}_2$ method and the two examples
in Section~\ref{examples} which use the group $\G = \mathbbm{Z}_8$
need a soft breaking of this type.

We point out that the systematic soft breaking of a symmetry
through all possible dimension-2 terms is not unusual in physics;
indeed,
it is always assumed in supersymmetric models.
We further point out that we are assuming the soft breaking of $\G$
to occur \emph{exclusively} through dimension-2 terms,
but not through dimension-3 terms;
indeed,
bare mass terms $m_{ab} \bar \nu_{Ra} C \bar \nu_{Rb}^T
+ \mbox{H.c.}$ might also break $\G$ softly,
but we assume them to remain absent,
even when $\G$ is softly broken by dimension-2 terms
in the scalar potential, lest the texture of $M_R$ be disrupted.

\section{Conclusions}
\label{concl}

In this paper we have suggested
flexible,
general,
and systematic methods for enforcing zeros
in arbitrary entries of the fermion mass matrices
by means of Abelian symmetries.
Though we have mainly concentrated on the lepton sector,
the methods,
as described in Section~\ref{method},
are also applicable in the quark sector,
and in Grand Unified Theories as well.
The general methods do not,
however,
necessarily lead to the simplest realization of each texture;
in this context,
``simple'' means either a small Abelian group $\G$,
a small number of scalar multiplets 
(Higgs doublets and SM gauge singlets),
or a scalar potential with few terms.

Using the lepton sector for definiteness,
we have identified two instances
where a simplification of the first method,
in which the Abelian group $\G$ is the direct product of
three groups $\G \left( D_{La} \right)$,
three $\G \left( \ell_{Ra} \right)$,
and three $\G \left( \nu_{Ra} \right)$,
is possible.
They are the following:
\begin{enumerate}
\item Either $M_\ell$ or $M_D$ is diagonal.
In this case,
we identify either $\G \left( \ell_{Ra} \right)$
or $\G \left( \nu_{Ra} \right)$,
respectively,
with $\G \left( D_{La} \right)$, 
for $a = 1,2,3$,
and then $\G$ has six Abelian factors instead of nine. 
If both $M_\ell$ and $M_D$ are diagonal,
then we make a triple identification
and we arrive at three factors in $\G$. 
\item The texture zeros in $M_\ell$ and $M_D$
are at the same positions.
In that case we identify $\G \left( \ell_{Ra} \right)$
with $\G \left( \nu_{Ra} \right)$ for $a = 1,2,3$,
and use a real representation of $\G$ for all the Higgs doublets.
\end{enumerate}
A third instance,
not discussed in this paper,
occurs if one has a Grand Unified Theory;
in that case the number of matter multiplets is reduced,
for instance to three in $SO(10)$,
and consequently the group $\G$ is smaller.

There is an alternative method
in which $\G$ is always $\mathbbm{Z}_{12} \times \mathbbm{Z}_2$.
When applying that method,
specific textures may allow the use of groups $\mathbbm{Z}_n$
with $n < 12$.

In many cases,
simplifications beyond the above procedures are possible;
unfortunately,
it is hard to provide general rules for those simplifications.
Still,
it is obvious that,
the larger is the number of zeros in the texture,
the smaller will be the number of scalar multiplets needed.

It will often prove necessary to introduce soft breaking
of the Abelian group $\G$ in the scalar potential,
through terms of dimension two,
in order to avoid dangerous Goldstone bosons.
Moreover,
the models constructed through our methods will usually display
many flavour-changing neutral Yukawa interactions,
since various Higgs doublets provide the different entries
of each mass matrix.
Still,
as our models are well defined and protected by symmetries,
these effects are calculable.

In summary,
the methods described in this paper
allow one to embed arbitrary fermion-mass-matrix textures
with texture zeros
into renormalizable field theories. 
This lends more credibility to this popular way
of constraining the mass matrices. 
The embedding is not unique,
indeed,
there is a large degree of arbitrariness in it.
Having chosen one embedding,
it is possible,
in principle,
to calculate the radiative corrections to the relations among the
masses and mixing angles which were achieved by assuming texture zeros.

\vspace{6mm}

\noindent \textbf{Acknowledgements}:
WG and ASJ thank K.R.S.\ Balaji for illuminating discussions.
ASJ also thanks H.\ Minakata for hospitality,
and the \textit{Japan Society for Promotion of Science} for support.
The work of LL has been supported
by the Portuguese \textit{Funda\c c\~ao para a Ci\^encia e a Tecnologia}
under the project CFIF--Plurianual.

\end{document}